\documentclass[twocolumn,eqsecnum,preprintnumbers,floatfix,prb,showpacs,aps,amsmath,amssymb]{revtex4}
\usepackage{epsfig,psfrag}
\usepackage{dcolumn}
\usepackage{bm}
\usepackage{graphicx}
\usepackage{color}
\newcommand{\gz}{\mathbb{Z}} 

\begin{document}

\title{Transient fluctuation relations for time-dependent particle transport}

\author{Alexander Altland,$^1$ Alessandro De Martino,$^1$ 
Reinhold Egger,$^2$ and Boris Narozhny$^{1,3}$}

\affiliation{ $^1$Institut f\"ur Theoretische Physik, Universit\"at zu K\"oln,
D-50973 K\"oln, Germany\\ $^2$Institut f\"ur Theoretische Physik, 
Heinrich-Heine-Universit\"at, D-40225 D\"usseldorf, Germany
\\ $^3$Institut f\"ur Theorie der Kondensierten Materie,
Universit\"at Karlsruhe, D-76128 Karlsruhe}

\date{\today}

\begin{abstract}
  We consider particle transport under the influence of time-varying
  driving forces, where fluctuation relations connect the 
statistics of pairs of 
   time reversed evolutions of physical observables. In many
  ``mesoscopic'' transport processes, the effective many-particle
  dynamics is dominantly classical, while the microscopic rates
  governing particle motion are of quantum-mechanical origin. We here
  employ the stochastic path integral approach as an optimal tool to
  probe the fluctuation statistics in such applications. Describing
  the classical limit of the Keldysh quantum nonequilibrium field
  theory, the stochastic path integral encapsulates the quantum origin
  of microscopic particle exchange rates.  Dynamically, it is
  equivalent to a transport master equation which is a formalism
  general enough to describe many applications of practical interest.
  We apply  the stochastic path integral to derive general functional 
  fluctuation relations for current flow induced by time-varying forces. We show
  that the successive measurement processes implied by this setup do
  not put the derivation of \textit{quantum} fluctuation relations in
  jeopardy. While in many cases the fluctuation relation for a full
  time-dependent current profile may contain excessive information, 
  we formulate a number of reduced relations, and demonstrate their
  application to mesoscopic transport. Examples include the
  distribution of transmitted charge, where we show that the
  derivation of a fluctuation relation requires the combined
  monitoring of the statistics of charge \textit{and} work.
\end{abstract}


\maketitle

\section{Introduction}\label{sec1}

Fluctuation relations (FRs) have recently emerged as a new powerful 
set of concepts in statistical physics.
\cite{phystoday,bochkov1,bochkov2,schutz,sevick,marconi,esposito}
They relate the stochastic fluctuations of systems 
far from equilibrium to their dissipative properties,
thereby generalizing the well-known fluctuation-dissipation 
theorem,\cite{ll} and provide ways to
quantify the degree of irreversibility of nonequilibrium processes. 
The possibility to formulate exact 
statements (sometimes referred to as ``fluctuation theorems'') 
about generic nonequilibrium systems -- 
both classical\cite{bochkov1,bochkov2,jar,crooks1,crooks2} and
quantum\cite{esposito,gaspard1,hanggi} -- has ignited
a burgeoning research activity. On the experimental side,  
first tests have already appeared, e.g., for soft matter\cite{exp1,expSeif1,exp1b} or
 mesoscopic systems.\cite{exp2,expSeif2,exp3,exp4}

The concept of FRs is frequently applied
to the statistics of variables of thermodynamic
significance, e.g., work, heat, or entropy.  This is exemplified by
the Crooks relation\cite{crooks1,crooks2}
\begin{equation}\label{crooks}
\frac{P(W)}{P_b(-W)} =  e^{\beta (W-\Delta F)} , 
\end{equation}
where $P(W)$ is the probability that an amount of work $W$ is done
on a system in a given driving protocol, i.e., when an
external time-dependent force $f_t$ is applied to the 
system during a time interval $t\in [-\tau,\tau]$.   
According to the Crooks relation, the ratio to the
probability $P_b(-W)$ of negative work done during the
``backward'' protocol, i.e., when the time-inverted force $f_{-t}$
acts on the system, is given by a Boltzmann-type factor,
where $\beta=T^{-1}$ is the inverse temperature ($k_B=1$ throughout) 
associated with the initial equilibrium state, and
$\Delta F$ is the thermodynamic free energy difference between
final and initial state.  Relations of this type may be applied to gain
access to thermodynamic data (e.g., $\Delta F$) from fluctuation statistics.
{}From Eq.~(\ref{crooks}) one obtains the celebrated 
Jarzynski equality,\cite{jar}
\begin{equation}\label{jarzyn}
\left\langle e^{-\beta W}\right\rangle = e^{-\beta \Delta F},
\end{equation}
where the average is over all process realizations under the 
force protocol $f_t$. On average $W\ge \Delta F$, which means that the  ``sum
rule'' \eqref{jarzyn} controls the cumulative
 weight of rare events. Relations of this type provide rigorous 
 bounds on the behavior of thermodynamic observables.  
 Fluctuations around these bounds have been analyzed 
 recently.\cite{grosberg,jar2,our}

 The application of external forces in many-particle systems generally
 leads to transport, with current flow and re-adjustment of particle
 concentrations. For time-dependent driving forces, FRs for the
 ensuing current profiles carry particularly rich information.  Here,
 the notion of ``transport'' is to be interpreted in a very general
 sense: it may refer to the changing number of individuals in a
 biological quasi-species model, to the electric current flow in a
 mesoscopic conductor, to the number of agents in a chemical reaction,
 etc.  Unlike with FRs for "global" (i.e., integrated over time)
 variables, the time-resolved information on transient current flow is
 stored in a time-dependent function $I=\{I_t\}$, and one needs to
 study \textit{functional} probability distributions $P[I]$, rather
 than functions like $P(W)$.  To be specific, transport through a
 ``system'' exchanging particles with $M$ ``reservoirs'' is described
 by currents $I_{\nu=1,\ldots, M}$ flowing out of the system, and
 $P[I]\equiv P[I_1,\dots,I_M]$.  Importantly, due to the discreteness
 of particle exchange with the reservoirs, the current flow enhances
 the noise level of the system. Far off thermal equilibrium, this
 ``shot noise'' often becomes the dominant source of fluctuations, and
the self-consistent description of the feedback
 cycle of currents generating noise and noise affecting the current
 flow becomes an important issue.

Given the present interest in time-dependent mesoscopic transport
phenomena, see for instance Refs.~\onlinecite{belzig,abanov,zhang}, or
in the work statistics under a quantum quench,\cite{silva} 
it is important to extend the general constraints imposed by 
transient FRs to the quantum setting. 
The above argument, however, needs to be applied with care in the quantum case.
In fact, the existing literature on FRs appears to
be essentially divided into a classical and a quantum part. This
concerns both the systems under consideration as well as
methodological aspects. Specifically, quantum theories of
FRs\cite{esposito,gaspard1,hanggi,gaspard2,hanggi2a,hanggi2b} mostly rely on quantum
measurement processes performed at the beginning and at the end of the
protocol.  This prescription is not directly suitable to
transient situations, where a continuous readout of, say, currents is
required. The lack of commutativity of current operators at different
times then becomes an issue, and the construction of a general quantum
theory of transient FRs may seem a difficult task.  Perhaps
surprisingly, the derivation of a FR for \textit{quantum} current 
flow goes through in unaltered form as long
as the external driving forces vary on time scales corresponding to classical
frequencies. We will discuss this point in some detail below
(see also Ref.~\onlinecite{hanggi2c}).
The situation becomes particularly transparent in the
many cases where the quantum system of interest actually operates
close to the \textit{semiclassical} limit: involving the dynamics of
many particles, the action scales relevant to transport are usually
much bigger than $\hbar$.  In this case, the quantum Keldysh
functional~\cite{alex,nazarov,kamenev_noneq,keldysh,schon,aleiner,ourprl} stays close to its
classical limit, the stochastic path integral,~\cite{kubo,metiu,hanggi1,pilgram}
and the formal lack of commutativity of current operators ceases to be an issue.

In an important early work,\cite{bochkov2} Bochkov and Kuzovlev (BK)
have formulated a general classical FR for current
flow. BK relied on a symmetry analysis of the Markovian
operators generating the dynamics of the system (see also the discussion in
Ref.~\onlinecite{hanggi2b}).  In contrast, our derivations below are
based on a path integral representation, which permits to explore the
nonequilibrium fluctuation statistics of observables beyond the
rigorous bounds imposed by FRs.  
Also, it stands to reason that the BK functional relation is too
general to be useful in applications. However, the general result can
be used to obtain more manageable \textit{derived fluctuation
relations}.  For example, rather than probing the full profile
$I=\{I_t\}$, one may consider the total charge transmitted into the
$\nu$th reservoir, $Q_\nu\equiv \int_{-\tau}^\tau dt\ I_{\nu,t}$. This
is arguably one of the most important global variables
characterizing a transport process.  Under \textit{stationary}
transport conditions, the forward and backward protocols coincide,
$P_b(Q)=P(Q)$, and a Crooks FR for charge has been stated in the
context of mesoscopic transport,\cite{esposito,tobiska}
\begin{equation}\label{crookscharge}
\frac{P(Q)}{P(-Q)} = e^{\beta \sum_\nu f_\nu Q_\nu}.
\end{equation}
This relation imposes nontrivial conditions on the generating
function for the full counting statistics (FCS)
of charge transport.\cite{fcs}  In particular, it implies
 that current cumulants of
different order must be linked 
together.\cite{tobiska,gaspard2,saito1,saito2,buttiker}
Equation (\ref{crookscharge}) has also been probed 
experimentally\cite{exp3,exp4} in mesoscopic circuits
using a quantum point contact charge detector.\cite{ensslin}
While the case of unidirectional single-electron counting 
 is directly accessible to experiments, recent progress has also 
 been reported for bidirectional counting.\cite{exp4,fujisawa} 
All these results apply to stationary regimes. Below we will show
that in the experimentally relevant case of \textit{time dependent}\
forces $f_t$ drastic things happen: the charge FR
(\ref{crookscharge}) actually breaks down, but a more general FR for
 the joint probability $P(Q,W)$ can still be formulated, see
 Eq.~(\ref{gencrooks}) below.  This ``generalized Crooks relation'' also
 ensures the validity of Eq.~\eqref{crooks} for the statistics of work
 alone. Furthermore, we will use the BK relation to derive 
cross-relations between
 nonlinear AC response coefficients, which can be put to an
 experimental test.

 The structure of the remainder of this article is as follows.  In
 Sec.~\ref{sec2} we discuss classical nonequilibrium transport
 processes   in terms of master equations and the stochastic path integral. 
 We derive the functional FR for currents under a transient driving protocol in
 Sec.~\ref{sec3}. The connection to the quantum theory is studied in 
 Sec.~\ref{sec4}.  In Sec.~\ref{sec5} we discuss derived fluctuation
 relations and compare them to numerical simulations for a mesoscopic
 RC circuit.  Some concluding remarks can be found in Sec.~\ref{sec6}.
 Various details of our calculations have been relegated to several Appendices. 

\section{Stochastic path integral}
\label{sec2}

\subsection{Master equation} \label{sec2a}

We are interested in the statistical properties of particle currents 
flowing through a system in contact with $M$ reservoirs. 
The probability $P_t(n)$ that the system contains $n$ 
particles at time $t$ is given by the convolution
\begin{equation} 
P_t(n) =\sum_{n_{-\tau}}P_t(n|n_{-\tau})\rho(n_{-\tau}),
\end{equation}
where $P_t(n|n_{-\tau})$ is the
conditional probability to evolve from an initial state $n_{-\tau}$ at
$-\tau$ to $n$ at $t$, and the weight $\rho(n_{-\tau})$ describes the
probability of the initial state. 
Without much loss of generality, we take $\rho$ to be of Boltzmann form,
\begin{equation} \label{eq:10}
 \rho(n) =  e^{-\beta (U(n)-F)},
\end{equation}
where $U(n)$ determines the internal energy of the system and
$F=-T\ln \sum_n \exp(-\beta U(n))$ is a free energy. 
The Markovian time evolution of $P_t$ is governed by a one-step master
equation,\cite{vanKampen}
\begin{eqnarray}\label{master}
\partial_t P_t(n) &=& -\hat
H_g P_t(n),\\ \nonumber
\hat H_g(n,\hat p)&=&\sum_{\nu=1}^M \sum_\pm  
\left(1-e^{\mp\hat p}\right) g^\pm_{\nu,t}(n),
\end{eqnarray} 
where the explicitly time-dependent 
rates $g_\nu^+$ ($g_\nu^-$) control the flux into (out of)
the system. The operator $e^{\hat p}$ 
($e^{-\hat p}$) raises (lowers) $n$ by one unit, i.e., we have 
the commutator $[\hat p,\hat n]=1$.
We assume that the rates obey the detailed balance condition
\begin{equation}\label{eq:1}
\frac {g^+_{\nu,t}}{ g^-_{\nu,t}} = e^{-\beta \kappa_{\nu,t}(n)},
\qquad \kappa_{\nu,t}(n)\equiv \partial_n U(n)-f_{\nu,t},
\end{equation}
where the functions $f_\nu$ describe external driving forces on the system.
For concreteness, we consider a cyclic protocol:
starting from an equilibrium situation at time
$t=-\tau$, $f_{\nu,-\tau}=0$, the time-dependent protocol $f_{\nu,t}\ne 0$ 
eventually ends at $f_{\nu,\tau}=0$.   
The generalization of the formalism below to several types of particles, 
or to multiple-step master equations is straightforward. 
However, the generalization to situations where the  reservoirs are at
different temperatures requires a more substantial extension of the
formalism. In this case, one has to account for the energy transfer 
necessarily accompanying particle transfer. We will briefly comment on
this point in Sec.~\ref{sec4c}.

The condition (\ref{eq:1}) is less restrictive than 
it might seem at first sight: it states that the
logarithmic ratio of rates is governed by a cost function, 
$T\ln(g^+_\nu/g^-_\nu) =-[ E_\nu(n+1)-E_\nu(n) ]$, measuring the
difference in ``energies'' $E_\nu(n)= -n f_\nu + U(n) $
before and after a particle has entered the system through terminal $\nu$. 
Note that $E_\nu(n)$ contains contributions
linear in the driving parameters and the particle number, i.e.,
the driving couples to the energy balance of individual particles and not
to particle interactions. 
With $U(n)$ introduced in Eq.~\eqref{eq:10} and the notation
$\partial_n U(n) \equiv U(n+1)-U(n)$, we obtain Eq.~\eqref{eq:1}. 
Particle interactions then correspond to nonlinearities in $U(n)$.

\begin{widetext}
\begin{table}[h]
 \centering
 \begin{tabular}{|c||c|c|c|}\hline
System&Variable $n$& $U(n)$& $f_{\nu,t}$\cr\hline\hline
electronic circuits\cite{nazarov}&charge&charging energy
&bias voltages\cr\hline
molecular motors\cite{KolomeiskyFisher}&mechanochemical state of motor
protein&load potential&ATP concentration\cr\hline
chemical reaction networks\cite{SchmiedlSeifert}& number of reaction
partners&internal energy&chemostat concentrations\cr\hline
adaptive evolution\cite{MustonenLassig}&allele frequencies&$\log$ equilibrium dist.&fitness gradients\cr\hline 
 \end{tabular}
 \caption{Examples of application fields for the master equation.}
 \label{tab:1}
\end{table}
\end{widetext}

Table \ref{tab:1} lists several application fields of present interest
where the above model of discrete transport applies with little or no
modification.  For later reference, let us introduce one of the
examples above in some more detail: consider a mesoscopic RC circuit,
where the system corresponds to a central node (``quantum dot'') with
$n$ electrons held by it.  The role of the reservoirs is taken by
$M=2$ voltage sources connected to the dot through resistors $R_1$ and
$R_2$.  The driving of the system by a time-varying bias voltage $V_t$
and its internal energy are given by $f_{\nu,t}= (-)^{\nu+1} 
eV_t/2 - eV_{\mathrm{eff}}$
(with $\nu=1,2$) and $U(n) = (n-1/2)^2 E_c$, respectively. Here,
$eV_\mathrm{eff}$ is the effective chemical potential 
on the dot that needs to be determined self-consistently, 
$E_c=e^2/(2C)$ is the capacitive charging energy of the dot, and we
have taken an offset charge corresponding to Coulomb blockade peak
conditions.\cite{nazarov} (In the rest of the paper we will set $e=1$.)
In earlier studies,\cite{exp1b,tc,zon1,zon2}
circuits of this type have been discussed within the framework of
Langevin equations, where the dominant source of fluctuations was
thermal noise.  In contrast, we wish to include the more general
mechanism of noise self-generated by transport.  This physics can be
described by the master equation \eqref{master} with sequential
tunneling rates,\cite{nazarov}
\begin{equation}\label{raterc}
g_{\nu,t}^\pm (n) = \frac{1}{R_\nu} \frac{\pm \kappa_{\nu,t}(n)}{e^{\pm\beta 
\kappa_{\nu,t}(n)}-1},
\end{equation}
with $\kappa_{\nu,t}$ in Eq.~\eqref{eq:1}.  The Bose-Einstein function 
in Eq.~(\ref{raterc}) indicates a degree of quantum-mechanical input,
to be discussed in more detail in Sec.~\ref{sec4c}.
The rates (\ref{raterc}) comply with the balance relation (\ref{eq:1}).
 In addition, it is straightforward to perform numerical simulations
for the dynamics generated by the master equation, see Sec.~\ref{sec5}.

\subsection{Path integral and current statistics} \label{sec2b}

To obtain the stochastic path-integral representation\cite{kubo,metiu}
of the above master equation, we allude to quantum-mechanical
analogies and notice the similarity of Eq.~\eqref{master} to an
(imaginary-time) Schr\"odinger equation.  The unit-normalized quantity
\begin{equation}
Z\equiv \sum_n P_\tau(n)=1
\end{equation} 
can thus be written as imaginary-time path integral,\cite{foot0}
\begin{equation}\label{pathin}
Z=\int D(n,p) \,
e^{\int_{-\tau}^\tau dt\left(p\partial_t n-H_g(n,p)\right)}\rho(n_{-\tau}).
\end{equation}
The integration in Eq.~(\ref{pathin}) is over smooth paths
 $(n,p)=\{(n_t,p_t)\}$, where
the auxiliary ``momentum'' $p_t\in i\mathbb{R}$ is integrated over the 
imaginary axis and the information on the actual discreteness of
the evolution of $n_t$ is encoded in the ``Hamiltonian'' $H_g$.

In order to extract information on current
flow, we need source fields. Specifically, time-dependent
counting fields $\chi_{\nu,t}$
probing the current flow $I_{\nu,t}$ can be introduced by
generalization of the Hamiltonian,\cite{naz}
\begin{equation}\label{hammast}
H_g(n,p)\to H_g(n,p,\chi)\equiv \sum_{\nu,\pm} 
\left(1-e^{\mp(p-i\chi_{\nu})}\right) 
g^\pm_{\nu}(n).
\end{equation}
Cumulants of the currents can then be obtained by functional differentiation,
\begin{eqnarray}\nonumber
 \langle\langle
I_{\nu_1,t_1}I_{\nu_2,t_2}\dots \rangle\rangle&=& \frac{i\delta}{
 \delta\chi_{\nu_1,t_1}}\frac{i\delta}{ \delta\chi_{\nu_2,t_2}}\dots
\Big|_{\chi=0} \ln Z[\chi] \\ \label{eq:14}
\Leftrightarrow
 Z[\chi]& =&\left\langle 
e^{-i \sum_\nu \int_{-\tau}^\tau dt\, \chi_\nu I_\nu}\right\rangle,
\end{eqnarray}
where $Z[\chi]$ is the generating functional, see Eq.~(\ref{eq:5}) 
below. Heuristically, the identification $\frac{i\delta}{\delta
 \chi} \leftrightarrow I$ follows from the fact that $\chi_\nu$ enters
the theory precisely like a vector potential. Much like in quantum
mechanics, differentiation with respect to (w.r.t.)
these vector potentials generates currents. More rigorously,
the connection follows from a ''Ward identity'' of
the functional integral: temporarily considering the case of identical
counting fields, $\chi_\nu= \chi$, the above
functional derivative obtains $\frac{i\delta}{\delta\chi_t} \ln Z=
\sum_\nu \langle I_{\nu,t}\rangle$, which we tentatively identify as
the total current out of the system. On the other hand, $\chi$ may be
gauged out of the Hamiltonian by a shift $p\to p+i \chi$, at the
expense of an extra term $i\int dt\, \chi \partial_t n$ in
the action. Differentiation w.r.t.~$\chi$ in the shifted
representation obtains $\frac{i\delta }{\delta\chi_t} \ln Z[\chi] =
-\langle \partial_t n_t\rangle$. The equality of the two representations
yields the continuity equation,
\begin{equation}\label{eq:4}
 \langle \partial_t n\rangle+\sum_\nu\langle I_\nu\rangle=0.
\end{equation}
This shows that the differentiation w.r.t.~$\chi$ obtains the total current,
while differentation w.r.t.~the fields $\chi_\nu$ yields currents through 
individual interfaces.  

The functional partition sum is
\begin{eqnarray}\label{eq:5}
 Z[\chi]&=& \int D(n,p) \, e^{-S_g[n,p,\chi]}\rho(n_{-\tau}),\\
\nonumber 
S_g[n,p,\chi]&=& - \int_{-\tau}^\tau dt\left(p\partial_t n-
H_g(n,p,\chi)\right),
\end{eqnarray}
with the Hamiltonian \eqref{hammast}.
According to standard rules of probability theory,
the probability distribution of currents
follows by functional integration over all $\chi$,
\begin{equation}\label{pf}
P[I] = \int D\chi \ e^{i\sum_\nu 
\int_{-\tau}^\tau dt\ \chi_{\nu} I_{\nu} } Z[\chi].
\end{equation}
The discreteness of particle transport is encoded in the exponential
dependence of the Hamiltonian on the ``phase space momentum''
$p$. Keeping this information is crucial, e.g., to properly
resolve the statistics of rare events.  
When the discreteness does not play an important role, an expansion of
$e^{\pm p}$ to quadratic order in $p$ may be justified. This reduces the 
stochastic path integral to the Martin-Siggia-Rose functional.\cite{msr,hkj,cd} 
One may then continue to either integrate over momenta, 
which leads to the Onsager-Machlup path integral,\cite{onsager,tc} or decouple 
the quadratic momentum dependence by an auxiliary ``noise field'', which
generates an effective Langevin description.\cite{kurchan,langevin2}
In the next section, we will employ the stochastic path integral \eqref{eq:5} to
(re)derive a number of general FRs.

\section{Time reversal and transient fluctuation relation for currents}
\label{sec3}

We proceed to derive a variant of the BK fluctuation 
relation\cite{bochkov2} for the statistics of
transient current flow. 
The properties of the stochastic path integral \eqref{eq:5}
rely on two fundamental symmetries, namely the continuity equation
\eqref{eq:4} and a symmetry under time reversal. The latter is
crucial to all FRs.  Our aim is thus to relate the functional $Z$,
describing evolution under the influence of rates $g=\{g_{\nu,t}\}$, 
to the functional $Z_b$ computed for the
time-reversed rates $(\hat T g)_t= g_{-t}$, i.e., 
$Z_b=Z\big|_{g\rightarrow  \hat T g}$.
Here we have defined a time reversal operator $\hat T$ that acts on 
``scalar'' functions $x=(n,g,f)$ as
$(\hat T x)_t= x_{-t}$, while ``vectorial'' functions $v=(I,p,\chi)$
transform as $(\hat T v)_t=-v_{-t}$.  As shown in Appendix \ref{appa}, 
the action in Eq.~\eqref{eq:5} satisfies the symmetry
\begin{eqnarray}\nonumber
 S_g[n,p,\chi]&=& S_{\hat T g}[\hat T n,\hat T (p-\beta 
 \partial_n U),\hat T(\chi+i\beta f)]\\ \label{eq:6} &+&
\beta [U(n_{\tau})-U(n_{-\tau})].
\end{eqnarray}
Next observe that $n$ and $p$ in Eq.~\eqref{eq:5} are just 
functional integration variables. With the auxiliary
relation for arbitrary functionals $F$,
\begin{eqnarray*}
 \int D(n,p) \,F[\hat T n,\hat T p] &=& \int D(\hat T n,\hat T p) 
\,F[\hat T n,\hat T p] \\ &=& \int D(n,p)\,F[n,p] ,
\end{eqnarray*}
substitution of Eq.~(\ref{eq:6}) into (\ref{eq:5}) 
and a shift of the momentum field, $p\to p +\beta \partial_nU$,
yields
\[
Z[\chi]= \int D(n,p) e^{-S_{\hat T g}[n,p,\hat T(\chi+i\beta
   f)]}\rho(n_{-\tau}).
\]
We thus obtain a prototypical FR for the generating functional, 
\begin{equation} \label{eq:7}
 Z[\chi] = Z_b[\hat T(\chi+i\beta f)].
\end{equation}

Inserting Eq.~(\ref{eq:7}) into \eqref{pf}, we arrive at a 
variant of the Crooks relation, first formulated by BK,\cite{bochkov2}
\begin{equation}  \label{eq:8}
 \frac{P[I]}{ P_b[\hat T I]} = e^{-\beta \int_{-\tau}^\tau dt\,
  \sum_\nu f_{\nu} I_{\nu}},
\end{equation}
where $P_b$ is the probability distribution
computed for time-inverted rates $\hat T g$.  
Integrating over $I$ and using  the normalization of $P_b[I]$, 
we also obtain a variant of the Jarzynski equality,
\begin{equation}
  \label{eq:3}
 \langle X \rangle = 1 ,\qquad X=\exp\left(\beta\int_{-\tau}^\tau
dt\sum_\nu f_\nu I_\nu\right).
\end{equation}
Equation \eqref{eq:8} represents the most general FR
relevant to this paper. Before turning to a discussion of its applications, 
it is worthwhile to link our present classical
formalism to the extended framework of a quantum theory of fluctuation
statistics. This will be the subject of Sec.~\ref{sec4}
below. Readers primarily interested in classical transport may skip
this section and directly turn to Sec.~\ref{sec5}. 
 
\section{Connection to Keldysh approach}
\label{sec4} 

In this section, we show how the stochastic path integral
(\ref{eq:5}) corresponds to the $\hbar\to 0$ limit of the Keldysh
quantum nonequilibrium functional. In this way, we will see how rates
encoding quantum statistics, cf.~Eq.~\eqref{raterc}, may appear as
dynamical input to an effectively classical theory of stochastic
fluctuations. To be concrete, we focus on the example of the
mesoscopic device introduced in section \ref{sec2a}, but generalization to
other setups is straightforward.

\subsection{Model}

We study a quantum dot connected to $M=2$ Fermi liquid leads
($\nu=1,2$) at temperatures well above the dot's mean level spacing
and in the ``open'' limit, $\bar g_\nu>1$. The dimensionless
conductances $\bar g_\nu\equiv 2\pi \hbar G_\nu$, with $G_\nu=R^{-1}_\nu$,
describe the
transparency of the contacts to the electrodes.\cite{nazarov,aleiner}
We assume that the external voltage $V_t$ varies on classical time
scales (such as the $RC$ time of the circuit), which are
large compared to quantum time scales of the problem.
Technically, this means that terms like $\hbar\partial_t V_t$ can be neglected.
The quantum nonequilibrium theory corresponds to a 
Keldysh functional integral,\cite{alex,nazarov,kamenev_noneq,keldysh} 
\begin{equation}\label{keld}
Z=\int D(\phi_c,\phi_q) \exp( -S[\phi_c,\phi_q] /\hbar ),
\end{equation}
with time-dependent ``classical'', $\phi_{c,t}$, and ``quantum'', $\phi_{q,t}$, 
phase fields.
These real-valued fields originate from a Hubbard-Stratonovich decoupling of 
the Coulomb interaction.\cite{alex}  
Note that in Eq.~(\ref{keld}), the action $S=S_c+S_{\mathrm{tun}}$
is dimensionful, in contrast to the dimensionless action
$S_g$ in the classical Eq.~\eqref{eq:5}. 
$S_c$ describes electron-electron interactions
(Coulomb blockade) due to the charging energy $E_c$, 
\begin{equation}\label{s0}
S_c[\phi_c,\phi_q] = i \frac{\hbar^2}{E_c} \int dt \ \phi_q  \partial_t^2\phi_c,
\end{equation}
while the tunnel action contains the influence of the attached electrodes
and is of Ambegaokar-Eckern-Sch\"on form,\cite{alex,schon,ourprl} 
\begin{equation}\label{stunnelact}
 S_{\rm tun}[\phi_c,\phi_q] = -\hbar \sum_{\nu=1,2} \frac{\bar g_\nu}{4} {\rm Tr} \left(
\hat \Lambda_\nu e^{-i\hat\phi} \hat \Lambda_{\rm d} e^{i\hat \phi} \right),
\end{equation}
where $\hat \Lambda_X$ (with $X=\nu,\mathrm{d}$) and $\hat \phi$ 
are operators in both Keldysh and time space.  We have
$\hat \phi=\{\phi_t\}$ with 
\begin{equation}\label{phidef}
\phi_t = {\rm diag}(\phi^+_t,\phi^-_t),\quad \phi^\pm=\phi_{c}\pm \phi_{q},
\end{equation}
where we employ the ``contour representation'' of Keldysh
theory\cite{keldysh}  throughout.
Furthermore, $\hat \Lambda_X =
\{\Lambda_{X,t,t'}\}$ with a $2\times 2$ matrix $\Lambda_{X,t,t'}$.
For classically varying $V_t$, it is convenient
to pass to a Wigner representation,\cite{alex} whereupon matrices become
functions of energy and time, $\{\Lambda_{X,t,t'}\}\to
\{\Lambda_{X}(t,\epsilon)\}$, and
 $\mathrm{Tr}\to (2\pi \hbar)^{-1} \int d\epsilon 
\int_{-\tau}^\tau dt\, \mathrm{tr}$, with
``tr'' denoting the trace in Keldysh space.
We then have the Keldysh matrix\cite{alex,nazarov,naz} 
\begin{equation}\label{lambda_c}
\Lambda_{X} (t,\epsilon) = \left( \begin{array}{cc} 1-2n_X & 2 n_X \\ 
2(1-n_X)  & -(1-2n_X)\end{array} \right),
\end{equation}
where $n_X=n_X(t,\epsilon)$ is the electron distribution
function in the leads ($X=\nu$) and in the dot ($X=\mathrm{d}$).
The Fermi-Dirac distribution function of the $\nu$th reservoir is
\begin{eqnarray}\label{fermifct}
n_\nu(t,\epsilon) &=& n_{\mathrm{F}}\left(\epsilon; 
V_{\nu,t},T\right) = \frac{1}{e^{(\epsilon-V_\nu)/T}+1},
\end{eqnarray} 
where the driving voltages are  
\begin{equation}\label{drivingvolt}
V_{\nu,t} \equiv (-)^{\nu+1} V_t/2.
\end{equation}
The dot distribution function $n_\mathrm{d}(t,\epsilon)$, however, has
to be determined self-consistently.  

Noting that the product between Wigner ``functions'' under the trace
in Eq.~\eqref{stunnelact} has to be understood as the Moyal product\cite{alex}
\[
A\ast B = A B + \frac{i\hbar}{2} (\partial_\epsilon A \partial_tB-\partial_t
A \partial_\epsilon B)+\mathcal{O}(\hbar^2),
\]
we obtain
\begin{equation}\label{auxx}
e^{-i\hat \phi}\hat\Lambda_{\mathrm{d}} e^{i\hat \phi}=
\left( \begin{array}{cc} 1-2n_{\mathrm{d},V_c}  
&2e^{-2i\phi_q} n_{\mathrm{d},V_c} \\ 2e^{2i\phi_q}(1-n_{\mathrm{d},V_c}) &
    -(1-2n_{\mathrm{d},V_c})\end{array} \right),
\end{equation}
where higher-order corrections in $\hbar$ are neglected, 
$n_{\mathrm{d},V_c}(\epsilon)\equiv n_\mathrm{d}(\epsilon-V_c)$, and
the dynamically fluctuating voltage on the dot is\cite{schon,ourprl} 
\begin{equation}\label{fluct}
V_{c,t} \equiv \hbar \partial_t \phi_{c,t}.
\end{equation}
Substituting Eq.~\eqref{auxx} into Eq.~\eqref{stunnelact}, we obtain
\begin{equation}\label{strewrite}
  S_\textrm{tun}[V_c,p]=\hbar 
\sum_{\nu,\pm}\int dt \left(1-e^{\mp p}\right)g_\nu^\pm(V_c)
\end{equation}
with $p=-2i\phi_q$ and the rates
\begin{eqnarray*}
  g_\nu^+(V_c)&=& G_\nu\int d\epsilon\, 
  n_{\nu}(\epsilon) [1-n_{\mathrm{d},V_c}(\epsilon)] , \\
  g_\nu^-(V_c)&=& G_\nu \int d\epsilon\, n_{\mathrm{d},V_c}(\epsilon)
  [1-n_{\nu}(\epsilon)].
\end{eqnarray*}
Finally, introducing a charge-like variable
through $n=V_c/2E_c$, adding the $(n,p)$ representation of the
charging term, $S_c[n,p]=-\hbar \int dt\ p\partial_t n$, and defining
$S_g[n,p]=S[n,p]/\hbar$, we arrive at an action as in Eq.~\eqref{eq:5},
where the Hamiltonian $H_g$ is governed by the rates
\begin{eqnarray} \label{rates_general}
  g_\nu^+(n)&= & G_\nu \int d\epsilon\, n_{\nu}(\epsilon + \partial_n U)
 \  [1- n_{\mathrm{d}}(\epsilon)],\\ \nonumber
  g_\nu^-(n)&= &G_\nu\int d\epsilon\, 
 [1- n_{\nu}(\epsilon+ \partial_n U) ] \ n_{\mathrm{d}}(\epsilon),
\end{eqnarray}
with $\partial_n U = 2E_c n$. 

\subsection{Model distributions} \label{sec4b}

Importantly, the rates \eqref{rates_general} do not determine the
action of the dot unless we specify the dot distribution function
$n_\mathrm{d}$. We here consider three different cases, all of which
are physically relevant and conceptually interesting in their own
way.  Which distribution is ultimately realized depends on the ratio of 
two time scales, the energy relaxation time due to electron-electron
interactions\cite{sivan,blanter,altshuler} on the dot,
$\tau_{ee}(\epsilon)=\hbar E_\mathrm{Th}^2/\epsilon^2 \Delta$, and
the time for escape into the leads, $\tau_{d}= \hbar /\Delta (\bar
g_1+\bar g_2)$. Here $\Delta$ is
the dot's single-particle level spacing, $\epsilon$ is the
characteristic excitation energy of particles in the system, and
$E_\mathrm{Th}=\hbar/t_\mathrm{Th}$ is the Thouless energy, 
where $t_\mathrm{Th}$ is the classical time scale before the
 single-particle dynamics in the dot becomes ergodic, e.g., 
the diffusion time.

The cases considered here are as follows.
(i) In the classical limit, $\hbar\to 0$, we observe that
$\tau_{ee}/\tau_d\to 0$. This implies
that strong relaxation mechanisms on the dot
enforce an effective equilibrium Fermi distribution,
\begin{equation}\label{n_hot}
n_\mathrm{d}(t,\epsilon)=n_{\mathrm{F}}(\epsilon;
V_{\mathrm{eff},t},T_{\mathrm{eff},t}),
\end{equation}
where the effective chemical potential $V_{\rm eff}$ and 
the effective temperature $T_{\rm eff}$ are determined by requiring particle
current and energy current conservation in the dot-leads composite
system,\cite{pilgram,pilgram2,nazarov} 
\begin{eqnarray}\nonumber
V_{\mathrm{eff},t} &=& \frac{\bar g_1 -\bar g_2}{\bar g_1+\bar g_2}
\frac{V_t}{2}, \\ \label{teff}
T_{\mathrm{eff},t} &=& 
\sqrt{T^2+  \frac{3 \bar g_1\bar g_2}{\pi^2 (\bar
g_1+\bar g_2)^2} V_t^2}.
\end{eqnarray}
This so-called ``hot electron distribution'' captures the heating of the system 
under the application of a voltage bias.  
We stress that Eq.~\eqref{n_hot} holds only when
$\tau_{ee}$ is short compared to the time scale for variation of $V_t$.
(ii) Alternatively, one may consider a situation where the system is
externally cooled to the ambient temperature $T$. In this case, we have 
\begin{equation} \label{n_cooled}
  n_\mathrm{d}(t,\epsilon)=n_{\mathrm{F}}(\epsilon;V_{\mathrm{eff},t},T).
\end{equation}
(iii) In cases where the dwell time $\tau_d$ is smaller than the
relaxation time $\tau_{ee}$, the effective distribution on the dot
is determined by the coupling to the leads rather than by internal
relaxation. In the absence of counting fields, $\chi=0$, this leads to 
the ``double-step distribution'' obtained by the weighted
superposition of the two lead distributions,
\begin{equation} \label{double_step}
  n_\mathrm{d}(t,\epsilon) =\sum_\nu \frac{\bar g_\nu}
{\bar g_1 + \bar g_2} n_\mathrm{F}(\epsilon; V_{\nu,t},T).
\end{equation}
However, for finite $\chi$, the situation gets more complicated in
that the effective dot distribution becomes $\chi$-dependent. The
ensuing structures are discussed in Appendix \ref{appb}, where
we also show that the FR \eqref{eq:7}, derived within a classical
formalism in Sec.~\ref{sec3}, stays valid in such a quantum-mechanical
setting.  

\subsection{Fluctuation relations} \label{sec4c}

In the remainder of this section, we address
the scenarios (i) and (ii) in some more detail.

First, if the dot is cooled down to ambient temperatures, the dot
distribution function is given by the Fermi
distribution \eqref{n_cooled}. Substituting this function into
Eq.~\eqref{rates_general} and doing the energy integrals, we obtain the
rates \eqref{raterc}. These rates obey the detailed balance relation
\eqref{eq:1}, which means that the externally cooled setup [case (ii)
above] seamlessly fits into the general framework of Secs.~\ref{sec2} and
\ref{sec3}.  Specifically, the FR \eqref{eq:8} and the derived 
relations in Sec.~\ref{sec5} below hold in full generality.

However, for a dot kept in isolation [case (i)], 
the situation is different. Here, the
temperature realized in the dot may differ strongly from that of the
leads, which means that there is no uniquely specified reference
temperature to relate to. Temperature mismatch of this type will 
effectively be realized in many different circumstances: out of 
equilibrium, transport through a dissipative system  generally 
leads to energy relaxation and, hence, to heating. 
It stands to reason that the effective
temperature will typically scale with the external driving parameters,
$T_\mathrm{eff}=\mathcal{O} (V_{\nu})$. The resulting effective
transport rates then no longer obey a detailed balance relation
containing the ambient temperature as a reference scale, which in turn
implies that the FR in Eq.~\eqref{eq:8} no longer holds.  

Albeit the FR \eqref{eq:8} is violated, it is still possible
to formulate modified FRs that contain crucial information about
the fluctuation statistics of the system.
Let us briefly sketch two different approaches to handling this 
situation. First, one may require that FRs categorically have to relate to
ambient temperature, or, more generally, to the temperatures $T_\nu$ 
of the external leads.  (In this part, 
we allow for unequal temperatures in the reservoirs.)
While our discussion above shows that the FR \eqref{eq:8} for particle
currents is violated, it is possible to derive joint FRs for particle 
and energy currents that do apply to the heated dot yet contain
the reservoir temperatures $T_\nu$ only. 
These FRs rely on the results for the statistics of charge and
 energy transfer  obtained in
 Refs.~\onlinecite{pilgram,pilgram2,kindermannpilgram,hekkilanazarov}.
For the convenience of the reader, we briefly
summarize the main conclusions of these references 
in the language of our paper. 
The main idea is to generalize the state space of the theory 
from $n$ to $(n,\epsilon)$, where
the continuous variable $\epsilon$ represents the energy of 
the dot. It is then straightforward to derive a generalized master equation
for the joint probability $P(n,\epsilon)$, where the rates
$g^\pm_{\nu}(n,\epsilon)$ 
are determined by the energy integrands of Eq.~\eqref{rates_general},
i.e., $g^\pm_\nu(n)=\int d\epsilon\, g^\pm_\nu(n,\epsilon)$. The
corresponding stochastic path integral is given 
by\cite{pilgram,kindermannpilgram,hekkilanazarov}
\begin{eqnarray}
Z &=& \int D(n,p,\epsilon,\xi) e^{-S_g[n,p,\epsilon,\xi]},  \\ \nonumber
S_g &=& - \int_{-\tau}^{\tau} dt \left[ p \partial_t n + \xi \partial_t 
\epsilon - H_g(n,p,\epsilon,\xi) \right], \\ \nonumber
H_g &=& \sum_{\nu,\pm} \int d\epsilon' 
\left(1-e^{\mp(p+\epsilon' \xi)}\right) g^{\pm}_\nu(n,\epsilon'), 
\end{eqnarray}
where the time-like variable $\xi$ is conjugate to $\epsilon$.
In order to probe the joint statistics of particle and 
energy currents,\cite{kindermannpilgram,hekkilanazarov} 
we couple both $p$ and $\xi$ to counting fields in the
respective $\nu$-dependent part of $H_g$,
\begin{equation}\label{newcounting}
p_t\to p_t - i \chi_{\nu,t},\quad \xi_t\to \xi_t-i \lambda_{\nu,t},
\end{equation}
see also Eq.~\eqref{hammast}.
One can then verify that the invariance of the action under
time reversal can be effected by a simultaneous 
transformation of both counting fields, 
\[
\chi_\nu \to \hat T(\chi_\nu + i \beta_\nu V_\nu),  \quad
\lambda_\nu \to \hat T(\lambda_\nu - i \beta_\nu),
\]
where $\beta_\nu\equiv T_\nu^{-1}$.
In this way, the symmetry relation in Eq.~\eqref{eq:6} 
gets generalized and we obtain the \textit{extended FR}
\begin{equation}\label{extfr}
Z[\chi_\nu,\lambda_\nu] = Z_b[ \hat T(\chi_\nu+i\beta_\nu V_\nu),
 \hat T(\lambda_\nu-i\beta_\nu)].
\end{equation}
Note that this FR involves the ambient (reservoir) temperatures
$T_\nu$ only. The price to be paid is that both the
particle ($\chi_\nu$) and the energy ($\lambda_\nu$) current
into the $\nu$th reservoir have to be monitored. 
 A quantum mechanical stationary version of Eq.~\eqref{extfr}
has recently been discussed in Ref.~\onlinecite{gaspard2}.

A second, and at this stage more heuristic, approach is 
to \textit{define} time-dependent effective temperatures $T^\ast_\nu$ 
characterizing the particle exchange with the $\nu$th reservoir 
through the logarithmic ratio of rates,
\begin{equation} \label{eq:2}
  T^\ast_\nu \equiv \frac{ V_\nu - V_{\mathrm{eff}} -\partial_n U }
  {\ln (g_\nu^+/g_\nu^- )}.
\end{equation}
We now observe that the FR in Eq.~\eqref{eq:7} holds
provided one replaces the global temperature $T=\beta^{-1}$ 
by the time- and $\nu$-dependent temperatures $T^{\ast}_{\nu}$
in Eq.~\eqref{eq:2}.   This  FR contains effective (and potentially
unknown) temperatures
different from the ambient temperature. Focusing on
particle transport alone,  it  provides a reduced 
description of the nonequilibrium process.  
Referring to the applications discussed in Sec.~\ref{sec5},
one may employ the statistical information encoded in such FRs
to \textit{determine} these temperatures. This is of
interest for systems where external biasing is expected to generate
heating through mechanisms that are not completely understood a priori. 
A precise formulation, however, is beyond the scope of this article
and requires to carefully address several subtleties.\cite{footeff}

In the next section, we will turn back to the general FR
\eqref{eq:8} and discuss its applied consequences in the description of
the fluctuation statistics of nonequilibrium transport. 

\section{Applications}
\label{sec5}

In the present approach, the general FR  \eqref{eq:8} and its spin-offs
are embedded into the formalism of the stochastic path integral. This
implies a lot of freedom in exploring the role of fluctuations
beyond the rigorous bounds implied by FRs. For
example, the general Jarzynski relation \eqref{eq:3} states that, for a cyclic protocol, 
the random variable $X$ averages to unity, $\langle X\rangle=1$. This
identity holds under very general circumstances, and it is in this
sense that the \textit{fluctuations} of $X$ contain more
telling information. This point has been explored in Ref.~\onlinecite{our},
where we showed how the statistics of $X$ signifies the crossover from near
into far equilibrium situations. 

In this paper, we concentrate on the statistical information
encoded in the FRs as such. A first aspect to notice
is that the \textit{functional} FR
\eqref{eq:8}, which is  a relation for the ``infinitely many'' variables
$I=\{I_t \}$, contains information that in most applications will be
excessive.  The applied value of the identity rather lies in its
potential as a starting point for the derivation of a wealth of
\textit{derived relations}.  Technically speaking, one may pass to
these reduced identities by taking marginals in
the sense of probability theory.  Below, we discuss
such reduction schemes on a number of examples. To keep the discussion
concrete, we will stay with our prototypical mesoscopic circuit as a
reference system.  Generalization to other systems should be 
straightforward.  We stress that the FR \eqref{eq:8} 
holds also in a quantum-mechanical setting, see our
discussion above and Ref.~\onlinecite{hanggi2c}.

\subsection{Stationary case}
 
Let us begin by discussing what Eqs.~(\ref{eq:7}) and (\ref{eq:8})
predict for the specific case of a stationary bias,
$V_t=V$. More precisely, we assume a symmetric protocol $V_t=V_{-t}$
which is switched on (off) within a time $\tau_s$ that is very short compared
to the counting time, $\tau_s\ll \tau$.  Then, $V_t$
assumes a constant value, $V$, during the long time interval 
$|t|<\tau-\tau_s$. For this voltage bias protocol, we have
$Z_b=Z$. Moreover, it makes sense to
consider the time averaged current, $I \equiv Q/2\tau$, where $Q=\int
dt I_t$ is the charge transmitted during the counting interval
$2\tau$.  Taking the limit $\tau_s/\tau\to 0$, a stationary bias 
can then be described using the above formalism.

Assuming a mesoscopic two-terminal ($M=2$) setup for
concreteness, let us choose constant counting fields, $\chi_1=- \chi_2
= \chi/2$, whereupon $Z[\chi]$ reduces to a \textit{function}
$Z(\chi)$, and differentiation w.r.t. $\chi$ probes
the statistics of $I$. The reduced form of Eq.~(\ref{eq:7}) then recovers
the known FR~\cite{esposito,tobiska,gaspard2}
\begin{equation}\label{ssr}
Z( \chi) = Z (-\chi+i\beta V),
\end{equation}
which is equivalent to Eq.~(\ref{crookscharge}) for the 
probability distribution function $P(Q)$.

It is worthwhile to discuss an important consequence that 
Eq.~\eqref{ssr} entails for taking the classical limit of nonequilibrium
quantum theories. Within the Keldysh approach to nonequilibrium
quantum dynamics it is customary to  associate the
classical limit with a quadratic expansion in the quantum field,
cf.~Ref.~\onlinecite{kamenev_noneq} for a discussion of this point.
However, the FR \eqref{ssr} implies that this
expansion cannot be valid, unless one operates in a near equilibrium
setting. To see this, notice that the counting field $\chi$ couples
additively to the quantum field $\phi_q$. Thus, a quadratic action in
$\phi_q$ implies a quadratic $S[\chi]$. Now, consider the most general
quadratic \textit{Ansatz},
$\ln Z(\chi) = 2\tau [- i\langle I\rangle \chi + C_2\chi^2]$, where we
noted that differentiation w.r.t. $\chi$ at $\chi=0$ yields the
average current, $\langle I\rangle$. Consistency with Eq.~\eqref{ssr} requires 
 $C_2=\langle I\rangle/(\beta V)$. For $\langle I \rangle \sim V$,
 this states that the fluctuations of $I$ (determined by the second
 order of the expansion in $\chi$) are thermal, $\mathrm{var}(I) \sim
 T$. Thus, a quadratic expansion in $\phi_q$ is not capable of
 describing nonequilibrium noise and cannot be valid in general. 

 More generally, Eq.~(\ref{ssr}) implies constraints for the nonlinear
 coefficients describing the response of the $k$th current cumulant
 $\langle \langle I^k \rangle\rangle$ to the voltage at order $V^l$.
 Since derivatives w.r.t.~$V$ can be traded for derivatives
 w.r.t.~$\chi$ in Eq.~(\ref{ssr}), different coefficients
 with the same order $k+l$ are connected.\cite{esposito,saito1,saito2}
 The fluctuation-dissipation theorem linking thermal Johnson-Nyquist noise to the linear
 conductance follows from the lowest-order equation in this hierarchy,
 $l+k=2$.  Higher-order relations with $l+k>2$ represent
 generalizations to the nonequilibrium.  We discuss the extension of
 these relations to the time-dependent situation in Sec.~\ref{sec5c}.

\subsection{Generalized Crooks relation}

We return now to a general setting with $M$ reservoirs and
time-dependent forces.  In applications, one is often interested in
the FCS of charge, $Q[I]= \int_{-\tau}^\tau dt\,I_t$, transmitted at
one of the terminals.   (We drop the reservoir index $\nu$ for
notational simplicity.) For stationary bias, the probability
distribution function $P(Q)$ obeys a FR, see Eq.~(\ref{crookscharge}),
and one may ask whether this relation extends to time-varying driving.
To answer this question, we study the statistics of both the
transmitted charge $Q[I]$ and the work done on the system, $W_g[I] =
\int_{-\tau}^\tau dt\, I_t f_t$ in terms of their joint probability density
\begin{eqnarray*}
 P(Q,W)&\equiv&\langle \delta(Q-Q[I]) \ \delta(W-W_g[I]) \rangle
\\  &=&\int  
\frac{d\chi_q d\chi_w}{ (2\pi)^2} e^{i (\chi_{q}
 Q + \chi_{w} W)} Z[\chi_q + \chi_w f].
\end{eqnarray*} 
The second equality follows from the integral representation of 
the $\delta$-functions, where, by virtue of Eq.~\eqref{eq:14}, 
the generating function $Z$ has to be taken for
the particular time-dependent counting field $\chi_t 
=\chi_q + \chi_w f_t$.
We now use the FR \eqref{eq:7} and integrate over the parameters
$\chi_{q,w}$,
\begin{align*}
&P(Q,W)= \\
&= \int 
\frac{d\chi_q d\chi_w}{ (2\pi)^2} e^{i (\chi_{q}
 Q + \chi_{w} W)}  Z_b[-\chi_q - (\chi_w -i\beta) {\hat T} f ] \\
&= 
\int \frac{d\chi_q d\chi_w}{ (2\pi)^2} e^{-i [\chi_{q} Q + (\chi_{w}+i\beta) W ]} 
Z_b[\chi_q + \chi_w {\hat T} f ] \\ 
&= e^{\beta W}
\left\langle \delta(Q+Q[I]) \ \delta(W+W_{\hat T g}[I])\right \rangle_b.
\end{align*}
Note that the transmitted charge and the work in the backward average 
are defined with the time-inverted force.  In effect, we obtain a 
generalized Crooks relation coupling
the transmitted charge and the work done on the system, 
\begin{equation} \label{gencrooks}
 \frac{P(Q,W)}{P_b(-Q,-W)} = e^{\beta W},
\end{equation}
which applies for time-dependent driving forces.
The derived transient FR (\ref{gencrooks}) 
connects the charge to the work fluctuation statistics.
Note that in a stationary situation, $W\propto Q$, and $P(Q,W)$
reduces to $P(Q)$. In that case, Eq.~(\ref{gencrooks}) 
implies the FR (\ref{crookscharge}) which now has the
same physical content as Eq.~(\ref{crooks}).  
Turning to the generic time-dependent case,
integrating Eq.~(\ref{gencrooks}) over $Q$ recovers the standard Crooks 
relation (\ref{crooks}) for the work distribution function $P(W)$.
However, there is no FR for the reduced probability
$P(Q)=\int dW P(Q,W)$ anymore, unless the driving force is 
time-independent. 

\vspace{1cm}

\begin{figure}[h]
 \centering
 \includegraphics[width=8cm]{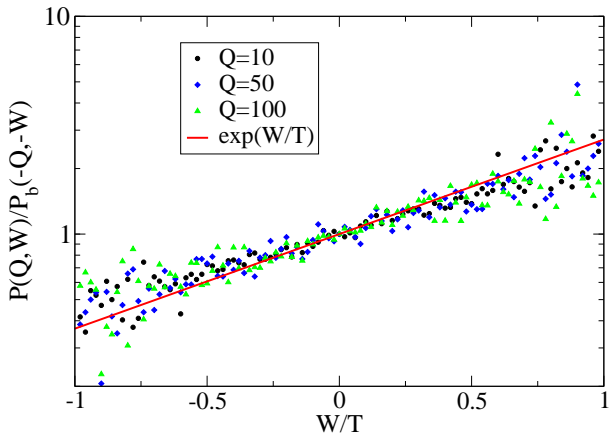}
 \caption{Numerical results for  $P(Q,W)/P_b(-Q,-W)$ vs $W/T$ for 
several values of the total transferred charge, taken within the 
respective window $[Q-1,Q+1]$.   The straight red line is the result 
predicted by Eq.~\eqref{gencrooks}. 
Parameters:  $\gamma=0.01\tau, \Omega\tau=2\times 10^3$, $V_0\tau=10$,
and $V_0/T=0.04$.}       \label{fig1}
\end{figure}

Figure \ref{fig1} shows a test of the generalized Crooks
relation \eqref{gencrooks} in a numerical simulation of the 
master equation \eqref{master}.
Here, we have considered the asymmetric pulse protocol
\begin{equation}\label{pulse}
V_t = V_0 \ \frac{\gamma t}{ t^2 + \gamma^2},
\end{equation} 
where $V_0$ is the pulse strength and $\gamma$ the
pulse width.  A relevant time scale is the 
RC time,  $\Omega^{-1}= RC/2$, where we choose symmetric contacts, $R_1=R_2=R$.
We mention in passing that $P_b(-Q,-W)=P(Q,-W)$ for an antisymmetric protocol
like Eq.~(\ref{pulse}). Within statistical errors,\cite{footmore}
the fluctuation statistics of $W$ under a fixed total transferred charge $Q$, 
obtained numerically from $4\times 10^6$ independent runs, 
agrees well with the prediction \eqref{gencrooks}.

\subsection{Nonlinear time-dependent transport coefficients} 
\label{sec5c}

Another way to reduce the information contained in the 
transient FR for currents, Eq.~\eqref{eq:7}, 
is to employ a series expansion of the $k$th current cumulant 
in terms of the driving forces $f_{\nu,t}$,
\begin{widetext}
\begin{equation} \label{onsager1}
\langle\langle I_{\nu_1,t_1}I_{\nu_2,t_2}\dots \rangle\rangle 
 = \sum_{l=0}^\infty \frac{1}{l!}
\sum_{\nu'_1,\dots, \nu'_l} \int_{-\tau}^\tau dt_1'\cdots dt'_l \
L^{(k,l)}_{\nu_1,\dots,\nu_k; \nu'_1,\dots,\nu'_l}(t_1,\dots,t_k;t'_1,\dots,
 t'_l) \  f_{\nu'_1,t'_1} \cdots f_{\nu'_l,t'_l}.
\end{equation}
Together with Eq.~(\ref{eq:14}),
this defines the time-dependent nonlinear Onsager response coefficients,
\begin{equation}\label{onsager2}
L^{(k,l)}_{\nu_1,\dots,\nu_k; \nu'_1, \dots,\nu'_l}
(t_1,\dots,t_k;t'_1,\dots, t'_l)=  \frac{i^k \delta^{(k+l)}}
{ \delta f_{\nu'_1,t'_1} \dots \delta f_{\nu'_l,t'_l}
\delta \chi_{\nu_1,t_1} \dots \delta \chi_{\nu_k,t_k}
} \ln Z[\chi]{\big\vert}_{\chi=f=0}.
\end{equation}
\end{widetext}
They are symmetric under arbitrary separate permutations 
of the indices $(\nu_i,t_i)$ and $(\nu'_i,t'_i)$.
Due to the normalization condition $Z[0]=1$, all coefficients with $k=0$
identically vanish. Moreover, for $l=0$, they represent equilibrium 
correlations.  Causality provides another constraint: the response of the 
system to the driving force $f_{\nu'_i,t'_i}$ is restricted 
to times $t\ge t'_i$. This implies that Eq.~(\ref{onsager2}) must 
vanish whenever there is at least one $t'_j>{\rm max} (t_1,\dots,t_k)$.

Using the FR (\ref{eq:7}), we next show how
 for a given order $k+l$, different coefficients
in Eq.~(\ref{onsager2}) are linked together.\cite{bochkov1} 
The general relation for these time-dependent quantities 
is rather lengthy and given in App.~\ref{appc}, see Eq.~\eqref{b1}. 
All relations resulting for the
four lowest orders ($k+l\le 4$) are also specified explicitly 
in App.~\ref{appc}.
In experimental applications, it is often more useful to 
probe Fourier coefficients, and  we provide here the
spectral decomposition of  the relations in App.~\ref{appc}.
We use the Fourier convention $L(t) = \frac{1}{2\tau}\sum_{n\in\gz} 
e^{-i\pi n t/\tau} L(n)$ (for all time arguments), 
where the series terminates at the desired accuracy level.  
Since the coefficients \eqref{onsager2} are real functions, 
we have $L(-n)=L^*(n)$.  Rather than stating the general formula, we  now
quote the result for the few lowest orders, $k+l=2,3,$ and 4.  

>From Eq.~(\ref{b2}), the \textit{second-order relations} are
\begin{eqnarray}\label{f2}
\mathrm{Im} \ L^{(2,0)}_{\nu_1,\nu_2}(n_1,n_2) &=&0, \\  
\nonumber
\mathrm{Re} \ L^{(1,1)}_{\nu_1;\nu'_1}(n_1;n'_1) &=&
 -\frac{\beta}{2}  L^{(2,0)}_{\nu_1,\nu'_1}(n_1,n'_1) .
\end{eqnarray}
Noting that $L^{(1,1)}$ describes the linear
dissipative response of currents to driving, while $L^{(2,0)}$ is a
measure of equilibrium fluctuations, we recognize the second relation in 
Eq.~(\ref{f2}) as a
spectral representation of the fluctuation-dissipation theorem.
Moreover, the symmetry of Eq.~\eqref{onsager2} under permutations of indices 
implies the Onsager-Casimir reciprocity relation
\begin{equation}
\mathrm{Re} \ L^{(1,1)}_{\nu_1;\nu'_1}(n_1;n'_1) = 
\mathrm{Re} \ L^{(1,1)}_{\nu'_1;\nu_1}(n'_1;n_1) .
\end{equation}
There is no condition on $\mathrm{Im}\ L^{(1,1)}$.
Next, from Eq.~(\ref{b3}) we find the \textit{third-order relations}
\begin{eqnarray}\label{f3}
\mathrm{Re} \ L^{(3,0)}_{\nu_1,\nu_2,\nu_3} &=&0, \\ \nonumber
\mathrm{Im} \ L^{(2,1)}_{\nu_1,\nu_2;\nu'_1} &=&
 \frac{\beta}{2} \mathrm{Im}\ L^{(3,0)}_{\nu_1,\nu_2,\nu'_1}, \\ \nonumber
 \mathrm{Re} \ L^{(1,2)}_{\nu_1;\nu'_1,\nu'_2}  &=&
 -\frac{\beta}{2} \mathrm{Re} \left ( L^{(2,1)}_{\nu_1,\nu'_1;\nu'_2}  + 
L^{(2,1)}_{\nu_1,\nu'_2;\nu'_1} \right).
\end{eqnarray}
For simplicity, we here suppress the Fourier indices $(n_i,n'_j)$, 
which can be easily restored from the respective reservoir 
indices $(\nu_i,\nu'_j)$.
The last relation in Eq.~(\ref{f3}) states how the leading nonlinear response
of the current connects to the linear order of the current noise.
Finally, in \textit{fourth order}, we get
\begin{widetext}
\begin{eqnarray}\nonumber 
\mathrm{Im} \ L^{(4,0)}_{\nu_1,\nu_2,\nu_3,\nu_4} &=&0, \qquad
\mathrm{Re} \ L^{(3,1)}_{\nu_1,\nu_2,\nu_3;\nu'_1} =
 -\frac{\beta}{2} L^{(4,0)}_{\nu_1,\nu_2,\nu_3,\nu'_1},\\ 
\label{f4}
\mathrm{Im}\ L^{(2,2)}_{\nu_1,\nu_2;\nu'_1,\nu'_2} &=& -\frac{\beta}{2}\
 \mathrm{Im} \left( L^{(3,1)}_{\nu_1,\nu_2,\nu'_1;\nu'_2} + 
L^{(3,1)}_{\nu_1,\nu_2,\nu'_2;\nu'_1} \right) ,\\ \nonumber
\mathrm{Re}\ L^{(1,3)}_{\nu_1;\nu'_1,\nu'_2,\nu'_3} &=&
-\frac{\beta^3}{2} L^{(4,0)}_{\nu_1,\nu'_1,\nu'_2,\nu'_3}  
-\frac{\beta^2}{2} \mathrm{Re}  \left(
L^{(3,1)}_{\nu_1,\nu'_2,\nu'_3;\nu'_1} + 
L^{(3,1)}_{\nu_1,\nu'_1,\nu'_3;\nu'_2}+
L^{(3,1)}_{\nu_1,\nu'_1,\nu'_2;\nu'_3}  \right)   \\ \nonumber
&-& \frac{\beta}{2} \mathrm{Re} \left(
L^{(2,2)}_{\nu_1,\nu'_1;\nu'_2,\nu'_3} + 
L^{(2,2)}_{\nu_1,\nu'_2;\nu'_1,\nu'_3} +
L^{(2,2)}_{\nu_1,\nu'_3;\nu'_1,\nu'_2}  \right) .
\end{eqnarray}
\end{widetext}
The utility of such relations in the characterization of stationary
transport has been emphasized before.\cite{esposito,gaspard2,saito1,saito2}
Here we have applied the concept of cross-relations to time-dependent 
coefficients. 
Such relations provide a hierarchy of benchmark criteria for
time-varying transport measurements or numerical simulations. 

\section{Conclusions}\label{sec6}

In recent years, fluctuation relations have been recognized as a
potent tool in the characterization of the fluctuation statistics of
nonequilibrium systems. Here, we have focused on the adaption of
this concept to the fluctuations of transient currents, the motivation
being that transport in response to time-varying
forces represents the perhaps most direct way of
probing the nonequilibrium physics of complex systems.  We advocated
the stochastic path integral as an optimal tool to describe the emerging
feedback mechanism of currents inducing noise which in turn
affects the statistics of currents. In its most general form, 
the degree of irreversibility by which a transport process
differs from the time-reversed process is characterized by the
functional Crooks relation \eqref{eq:8}. This relation has first been
stated in the seminal BK paper.\cite{bochkov2} The value of the present
derivation primarily lies in the linkage of the exact ``sum
rule''~\eqref{eq:8} to the highly flexible formalism of the stochastic
path integral. Indeed, we argued that the fluctuation relation
is overly general to be of much use in concrete applications. It is
more informative to explore fluctuations around the rigorous bounds
imposed by these relations, or to consider derived relations for the
statistics of time integrated variables, or Fourier coefficients. 

The stochastic path integral also affords an interpretation as the
classical limit of the quantum theory of nonequilibrium
fluctuations. This connection is useful both from a conceptual and an
applied point of view. Conceptually, it provides a connection between the
quantum and the classical theory of fluctuation relations. The former
is usually described by means of quantum projector techniques, 
a theoretical language that is not straightforwardly linked to classical limits.
>From an applied point of view, the quantum-classical correspondence
is of value in that it allows for an integrated description of
processes whose stochastic rates follow from a microscopic quantum
dynamics.  Moreover, we have demonstrated that the FR stays valid
in a quantum-mechanical regime.

In this paper, we illustrated many of the concepts above on the
prototypical example of a mesoscopic RC circuit. However, the
underlying theoretical framework is much more general in nature, and
we believe that it can be straightforwardly adjusted to many other different settings. 

\section{Acknowledgements}

This work was supported by the SFB TR 12 of the DFG and by the
Humboldt foundation.

\vspace{1cm}

\appendix

\section{Symmetry of the action}\label{appa}

Here we provide the derivation of the symmetry relation (\ref{eq:6})
for the action \eqref{eq:5} of the stochastic path integral.
Using the definition of the time-reversal operator $\hat T$ in 
Sec.~\ref{sec3} 
and recalling the definition of $\kappa_{\nu,t}(n)$ in Eq.~(\ref{eq:1}),
we have 
\begin{widetext}
\begin{eqnarray*}
S_{\hat T g} [\hat T n , \hat T (p - \beta \partial_n U),\hat T( \chi
+ i\beta f)]  &=&  - \int_{-\tau}^\tau dt \Bigl[  p_{-t} \partial_{-t} n_{-t}+
\beta (\partial_n U)_{-t} \partial_{-t} n_{-t} \\ &-&
\sum_{\nu,\pm} \left( 1-e^{\mp[-p_{-t}
 + i \chi_{\nu,-t} -\beta \kappa_{\nu,-t}(n_{-t}) ]} \right) g_{\nu,-t}^\pm
\Bigr ]. 
\end{eqnarray*}
Next we change the time integration variable, $t\to -t$, and
the summation variable $\pm\to \mp$.
Noting that $\int_{-\tau}^\tau dt \ (\partial_n U) \partial_t n = 
U(n_\tau)-U(n_{-\tau})$,
we obtain for the above expression
\[
S_{\hat T g} = -\beta [U(n_\tau)-U(n_{-\tau})] 
- \int_{-\tau}^\tau dt\ \left [ p_t \partial_t n_t  
- \sum_{\nu,\pm}  \left( 1- e^{\mp(p_t-i\chi_{\nu,t}+
 \beta \kappa_{\nu,t} (n_t)} \right) g_{\nu,t}^\mp \right].
\]
\end{widetext}
Finally, we employ the detailed balance condition 
\eqref{eq:1} and arrive at Eq.~\eqref{eq:6}.

\section{Double-step distribution}\label{appb}

In this appendix,\cite{foot0} we discuss the structure of the 
tunnel action (\ref{stunnelact}) when the effective dot distribution has the
double-step form in Eq.~\eqref{double_step}.
We assume that the system is driven and monitored on classical time 
scales such that we can neglect terms like $\hbar \partial_t V_\nu$ and
$\hbar \partial_t\chi_{\nu}$.  The driving voltages $V_\nu$ were defined in 
Eq.~\eqref{drivingvolt}.  In the present quantum-mechanical context,
it is crucial to include the counting fields $\chi_\nu$ from the outset.
The Wigner representation of the $\Lambda_X$ matrices 
generalizing Eq.~\eqref{lambda_c} is given by\cite{naz} 
\begin{eqnarray}\nonumber
\tilde \Lambda_\nu (t,\epsilon) &=& \left( \begin{array}{cc}
1-2n_\nu  & 2e^{i\chi_{\nu,t}} n_\nu  \\
2e^{- i\chi_{\nu,t}} (1-n_\nu ) & -(1-2n_\nu) 
\end{array} \right), \\ \label{lambdaDS} 
\hat\Lambda_\mathrm{d} &=& {\cal N} 
\sum_\nu \frac{\bar g_\nu}{\bar g_1 + \bar g_2}
 \tilde \Lambda_\nu  ,
\end{eqnarray}
where $n_\nu(t,\epsilon)$ is defined in Eq.~\eqref{fermifct}.
The factor ${\cal N}(t,\epsilon)$ ensures 
the normalization condition,\cite{alex} $\hat \Lambda_\mathrm{d}^2=1$, 
and is given by
\begin{eqnarray*}
{\cal N} &=&  \frac{\bar g_1+\bar g_2}{\sqrt{(\bar g_1 + \bar g_2)^2+ 4
 \bar g_1 \bar g_2 B_\chi}},  \\
B_\chi  &= & (e^{  i \chi } -1) n_1 (1-n_2)  + 
(e^{  -i \chi } -1) n_2 (1-n_1), 
\end{eqnarray*}
where $\chi \equiv \chi_1-\chi_2$. Note 
the symmetry property $B_{-\chi+i\beta V}=B_\chi$,
which is instrumental to prove FRs.  
Keeping terms up to order $\hbar \partial_t \phi_q$,
instead of Eq.~\eqref{auxx}, the Moyal product expansion yields
\begin{eqnarray}\label{declambda}
&& e^{i\hat  \phi} \tilde \Lambda_\nu e^{-i\hat \phi}= \\ && \nonumber
\left( \begin{array}{cc}
1-2n_{\nu,V_c+\hbar \partial_t \phi_q} 
 &2 e^{i(\chi_{\nu}+2 \phi_{q})} n_{\nu,V_c}  \\
2e^{- i(\chi_{\nu}+2\phi_q)}  (1-n_{\nu,V_c} ) & 
-(1-2n_{\nu, V_c - \hbar \partial_t \phi_q} ) 
\end{array} \right),
\end{eqnarray}
where $n_{\nu,V_c}(\epsilon)\equiv n_\nu(\epsilon+V_c)$. 
Notice that the energy arguments in the diagonal elements are shifted by
$\pm \hbar \partial_t \phi_q$.
We now insert Eqs.~(\ref{lambdaDS}) and (\ref{declambda}) 
into Eq.~\eqref{strewrite}, and write
$S_\mathrm{tun} = S_\mathrm{tun}^{(1)} + S_\mathrm{tun}^{(2)}+
S_\mathrm{tun}^{(3)}.$
The first term reads 
\begin{eqnarray} \nonumber
S^{(1)}_\mathrm{tun} &= & \frac{\hbar}{2} \sum_{\nu\nu',\pm} \int dt
\Bigl [ -2 e^{\pm i(2\phi_q + \chi_\nu-\chi_{\nu'})} g^\pm_{\nu \nu'}(V_c) 
\\  &+ & \label{stunn1} g^+_{\nu \nu'}(V_c \pm \hbar \partial_t\phi_q) 
+ g^-_{\nu \nu'}(V_c \pm \hbar \partial_t \phi_q) \Bigr], 
\end{eqnarray}
with the rates 
\begin{eqnarray*}
g^+_{\nu\nu'}(V_c) &= & 
\frac{G_\nu G_{\nu'}}{G_1+G_2} \int d\epsilon \ n_{\nu,V_c}(\epsilon)
\ [1-n_{\nu'}(\epsilon)],\\
g^-_{\nu\nu'}(V_c) &= &\frac{G_\nu G_{\nu'}}{G_1+G_2} 
 \int d\epsilon \  [1-n_{\nu,V_c}(\epsilon)] \ n_{\nu'}(\epsilon).
\end{eqnarray*}
These rates have the same functional form as the 
sequential tunneling rates in Eq.~(\ref{raterc}).
If $\phi_q$ fluctuates on classical time scales,
 i.e., $\hbar \partial_t \phi_q  \approx 0$, we see that
$S^{(1)}_\mathrm{tun}$ has the same structure as 
the Hamiltonian $H_g$ in Eq.~(\ref{hammast}).
Furthermore,
$S^{(2)}_\mathrm{tun}$ coincides with $S^{(1)}_\mathrm{tun}$, 
except that the rates $g^\pm_{\nu \nu'}$ are replaced by the modified rates 
\begin{eqnarray*}
\tilde g^+_{\nu\nu'}(V_c) &= & 
\frac{G_\nu G_{\nu'}}{G_1+G_2} \int d\epsilon \ ({\cal N}(t,\epsilon) -1)
\ n_{\nu,V_c} [1-n_{\nu'}],\\
\tilde g^-_{\nu\nu'}(V_c) &= &\frac{G_\nu G_{\nu'}}{G_1+G_2} 
 \int d\epsilon \   ({\cal N}(t,\epsilon) -1) \ [1-n_{\nu,V_c}] n_{\nu'}, 
\end{eqnarray*}
which are now complicated functions of the dynamical 
voltage $V_{c}$, see Eq.~\eqref{fluct}, of the bias voltage $V$,
and, via the normalization factor $\cal N$, of the 
counting field $\chi$.  These rates, and thus
 $S^{(2)}_\mathrm{tun}$, vanish in the absence of 
counting fields, since then ${\cal N}=1$.
It is easy to check that they still satisfy the crucial 
detailed balance condition \eqref{eq:1}, since
\[
\frac{\tilde g^+_{\nu \nu'}(V_c)}{\tilde g^-_{\nu\nu'}(V_c)} =
 e^{-\beta (V_c - V_\nu + V_{\nu'} )}.
\]
Finally, $S_\mathrm{tun}^{(3)} = -\hbar (G_1+G_2)
\int dt  d\epsilon \ ({\cal N}(t,\epsilon)-1)/2,$
which does not contribute to the field dynamics
but has to be retained for calculating the current cumulants. 

With the fields $\phi^\pm$ in Eq.~\eqref{phidef}, we now observe that
$S_{\mathrm{tun}}$ satisfies the symmetry property
\begin{eqnarray}\label{newsym}
&& S_{\mathrm{tun}}\left [\phi^+_{t},\phi^-_{t},\chi_{\nu,t}\right]
 = \\ \nonumber &&
S_{\mathrm{tun}}\left [-\phi^+_{-t + i \hbar \beta /2},
-\phi^-_{- t - i \hbar \beta /2}, - \chi_{\nu,-t} + i\beta V_{\nu,-t} \right],
\end{eqnarray}
which expresses the time reversal invariance of $S_\mathrm{tun}$.
This relation, together with the invariance of $S_c$ in Eq.~(\ref{s0})
under the replacement $\phi^\pm_t \to -\phi^\pm_{-t \pm  i \hbar \beta /2}$,
when inserted in the Keldysh generating functional (\ref{keld}),
leads to the quantum generalization  of the classical FR \eqref{eq:7}.
We notice that in the limit $\hbar\to 0$, 
the replacement $\phi^\pm_t \to -\phi^\pm_{-t \pm  i \hbar \beta /2}$  
in Eq.~\eqref{newsym} is equivalently written as 
\begin{eqnarray*}
\phi_{q,t} &\to&  -\phi_{q,-t} +i \beta V_{c,-t}/2 + {\cal O}(\hbar),\\
V_{c,t} & \to &  V_{c,-t} + {\cal O}(\hbar),
\end{eqnarray*}
where $\phi_q=(\phi^+-\phi^-)/2$ and
$V_c=\hbar\partial_t (\phi^++\phi^-)/2$. 
Equation (\ref{newsym}) thus recovers the classical relation \eqref{eq:6}
but allows us to extend the validity of the general 
FR \eqref{eq:7}, which we obtained in Sec.~\ref{sec3} from the classical 
generating functional, to the more general setting of the Keldysh 
quantum generating functional.

\section{Relations between time-dependent response coefficients}
\label{appc}

In this appendix, we provide some details concerning 
Sec.~\ref{sec5c}.  The general FR \eqref{eq:7}
for currents under time-dependent driving forces $f_{\nu,t}$
implies that derivatives w.r.t.~forces can be exchanged for
derivatives w.r.t.~counting fields.  The time-dependent Onsager 
coefficients (\ref{onsager2}) can thereby be written as 
\begin{widetext}
\[
L^{(k,l)}_{\nu_1,\dots,\nu_k; \nu'_1, \dots,\nu'_l}
(t_1,\dots,t_k;t'_1,\dots,t'_l) = \prod_{s=1}^k  
\left( \frac{\delta}{i \delta \chi_{\nu_s,-t_s}}  \right)
 \prod_{j=1}^l \left( i\beta \frac{\delta}{\delta 
\chi_{\nu'_j,-t'_j}} + \frac{\delta}{\delta f_{\nu'_j,-t'_j}} \right) 
 \ln Z[\chi]{\big\vert}_{\chi=f=0} .
\]
Therefore we obtain a connection between coefficients with 
the same order $k+l$,
\begin{eqnarray}\label{b1}
&& (-1)^k L^{(k,l)}_{\nu_1,\dots,\nu_k; \nu'_1, \dots,\nu'_l}
(t_1,\dots,t_k;t'_1,\dots,t'_l)
= \beta^l L^{(k+l,0)}_{\nu_1,\dots,\nu_k, \nu'_{1}, 
\dots,\nu'_{l} }(-t_1,\dots,-t'_l) \\ \nonumber
&& + \ \beta^{l-1} \sum_{j=1}^l 
L^{(k+l-1,1)}_{\nu_1,\dots,\nu_k,\nu_1,\dots, \bar \nu'_j,\dots \nu'_l;
 \nu'_j} (-t_1,\dots,-\bar t'_j,\dots, -t'_l; -t'_j) \\  \nonumber
&& + \  \beta^{l-2} \sum_{j\ne m=1}^{l} L^{(k+l-2,2)}_
{ \nu_1, \dots, \nu_k, \nu'_1, \dots, \bar \nu'_j, \dots, \bar \nu'_m,
\dots, \nu'_l ; \nu'_j, \nu'_m}  (-t_1,\dots,-\bar t'_j, 
\dots, -\bar t'_m,\dots, -t'_l; -t'_j, -t'_m) \\ \nonumber
&& + \cdots + L^{(k,l)}_{\nu_1,\dots,\nu_k; \nu'_1, \dots , 
\nu'_l}(-t_1,\dots,-t_k;-t'_1,\dots, -t'_l),
\end{eqnarray}
where $\bar \nu'_j$ (resp. $\bar t'_j$) means that $\nu'_j$ (resp. $t'_j$)
is missing in the string of indices (resp. time arguments).

For instance, the four lowest-order relations resulting 
from Eq.~(\ref{b1}) are as follows: 
To first order, $L^{(1,0)}_{\nu}(t)=-L^{(1,0)}_{\nu} (-t).$
The second-order result is 
\begin{eqnarray}\label{b2}
 L^{(2,0)}_{\nu_1,\nu_2}(t_1,t_2) &=&  L^{(2,0)}_{\nu_1,\nu_2}(-t_1,-t_2) 
\equiv \hat T L^{(2,0)}_{\nu_1,\nu_2} , 
\\ \nonumber L^{(1,1)}_{\nu_1;\nu'_1}(t_1;t'_1)  
&=& - \beta L^{(2,0)}_{\nu_1,\nu'_1}(-t_1,-t'_1)
 - L^{(1,1)}_{\nu_1;\nu'_1}(-t_1,-t'_1) = 
- \beta \hat T L^{(2,0)}_{\nu_1,\nu'_1}-\hat T L^{(1,1)}_{\nu_1;\nu'_1},
\end{eqnarray}
where the time reversal operator $\hat T$ inverts all time arguments 
when acting on a function. For $k+l=3$, we find the relations 
\begin{eqnarray}\label{b3}
L^{(3,0)}_{\nu_1,\nu_2,\nu_3} &= & - \hat T L^{(3,0)}_{\nu_1,\nu_2,\nu_3} , 
\qquad   L^{(2,1)}_{\nu_1,\nu_2;\nu'_1} =
\beta \hat T L^{(3,0)}_{\nu_1,\nu_2,\nu'_1} +
\hat T L^{(2,1)}_{\nu_1,\nu_2;\nu'_1}, \\
\nonumber L^{(1,2)}_{\nu_1;\nu'_1,\nu'_2}&=&
- \beta^2  \hat T L^{(3,0)}_{\nu_1,\nu'_1,\nu'_2} -
\beta \hat T \left(  L^{(2,1)}_{\nu_1,\nu'_2;\nu'_1}+
 L^{(2,1)}_{\nu_1,\nu'_1;\nu'_2} \right)  -
\hat T L^{(1,2)}_{\nu_1;\nu'_1\nu'_2} .
\end{eqnarray}
Finally, the fourth order produces the following relations:
\begin{eqnarray}\label{b4}
L^{(4,0)}_{\nu_1,\nu_2,\nu_3,\nu_4}& =&
 \hat T L^{(4,0)}_{\nu_1,\nu_2,\nu_3,\nu_4},\qquad
L^{(3,1)}_{\nu_1,\nu_2,\nu_3;\nu'_1} =
- \beta \hat T L^{(4,0)}_{\nu_1,\nu_2,\nu_3,\nu'_1} -
\hat T L^{(3,1)}_{\nu_1,\nu_2,\nu_3;\nu'_1}, \\ \nonumber
L^{(2,2)}_{\nu_1,\nu_2;\nu'_1,\nu'_2} &=&
\beta^2   \hat T L^{(4,0)}_{\nu_1,\nu_2,\nu'_1,\nu'_2}
+  \beta\hat T \left( L^{(3,1)}_{\nu_1,\nu_2,\nu'_2;\nu'_1}+
L^{(3,1)}_{\nu_1,\nu_2,\nu'_1;\nu'_2} \right) +
\hat T L^{(2,2)} _{\nu_1,\nu_2;\nu'_1,\nu'_2} , \\ \nonumber
L^{(1,3)}_{\nu_1;\nu'_1,\nu'_2,\nu'_3} &=&
-\beta^3  \hat T L^{(4,0)}_{\nu_1,\nu'_1,\nu'_2,\nu'_3} -  \beta^2
\hat T \left(L^{(3,1)}_{\nu_1,\nu'_2,\nu'_3;\nu'_1} +
L^{(3,1)}_{\nu_1,\nu'_1,\nu'_3;\nu'_2}+
L^{(3,1)}_{\nu_1,\nu'_1,\nu'_2;\nu'_3} \right)   \\ \nonumber
&& - \ \beta \hat T\left(L^{(2,2)}_{\nu_1,\nu'_1;\nu'_2,\nu'_3} +
L^{(2,2)}_{\nu_1,\nu'_2;\nu'_1,\nu'_3}+
L^{(2,2)}_{\nu_1,\nu'_3;\nu'_1,\nu'_2} \right) - 
\hat T L^{(1,3)}_{\nu_1;\nu'_1,\nu'_2,\nu'_3}.
\end{eqnarray}
\end{widetext}


\begin{thebibliography}{10}

\bibitem{phystoday}
C. Bustamante, J. Liphardt and F. Ritort, Physics Today {\bf 58}, 43 (2005).

\bibitem{bochkov1}
G.N. Bochkov and Yu.E. Kuzovlev, 
Zh. Eksp. Teor. Fiz. {\bf 76}, 1071 (1979)
[Sov. Phys. JETP {\bf 49}, 543 (1979)].

\bibitem{bochkov2}
G.N. Bochkov and Yu.E. Kuzovlev, 
Physica {\bf 106A}, 443 (1981).

\bibitem{schutz}
R.J. Harris and G. Sch\"utz, J. Stat. Mech. P07020 (2007).

\bibitem{sevick}
E.M. Sevick, R. Prabhakar, S.R. Williams, and D.J. Searles,
Annu. Rev. Phys. Chem. {\bf 59}, 603 (2008).

\bibitem{marconi}
U.M.B. Marconi, A. Puglisi, L. Rondoni, and A. Vulpiani,
Phys. Rep. {\bf 461}, 111 (2008).

\bibitem{esposito}
M. Esposito, U. Harbola, and S. Mukamel,
Rev. Mod. Phys. {\bf 81}, 1665 (2009). 

\bibitem{ll}
L.D. Landau and E.M. Lifshitz, \textit{Statistical Physics, Part 1}, 3rd edition
(Elsevier Butterworth-Heinemann, 1980).

\bibitem{jar}
C. Jarzynski, Phys. Rev. Lett. {\bf 78}, 2690 (1997).

\bibitem{crooks1}
G.E. Crooks, Phys. Rev. E {\bf 60}, 2721 (1999).

\bibitem{crooks2}
G.E. Crooks, Phys. Rev. E {\bf 61}, 2361 (2000).

\bibitem{gaspard1}
D. Andrieux and P. Gaspard, Phys. Rev. Lett. {\bf 100}, 230404 (2008).

\bibitem{hanggi}
M. Campisi, P. Talkner, and P. H\"anggi, 
Phys. Rev. Lett. {\bf 102}, 210401 (2009).

\bibitem{exp1}
D.M. Carberry \textit{et al.}, Phys. Rev. Lett. {\bf 92}, 140601 (2004).

\bibitem{expSeif1}
V. Blickle, T. Speck, L. Helden, U. Seifert, and C Bechinger,
Phys. Rev. Lett. {\bf 96}, 070603 (2006).

\bibitem{exp1b}
J.R. Gomez-Solano, A. Petrosyan, S. Ciliberto, R. Chetrite, and K. Gawedzki,
Phys. Rev. Lett. {\bf 103}, 040601 (2009).

\bibitem{exp2}
N. Garnier and S. Ciliberto, Phys. Rev. E {\bf 71}, 060101 (2005).

\bibitem{expSeif2}
S. Schuler, T. Speck, C. Tietz, J. Wrachtrup, and U. Seifert,
Phys. Rev. Lett. {\bf 94}, 180602 (2005).

\bibitem{exp3}
S. Nakamura \textit{et al.}, 
Phys. Rev. Lett. {\bf 104}, 080602 (2010).

\bibitem{exp4}
Y. Utsumi \textit{et al.}, Phys. Rev. B {\bf 81}, 125331 (2010).

\bibitem{grosberg}
R.C. Lua and A.Y. Grosberg, J. Phys. Chem. B {\bf 109}, 6805  (2005);
G.E. Crooks and C. Jarzynski, Phys. Rev. E {\bf 75}, 021116 (2007).

\bibitem{jar2}
C. Jarzynski, Phys. Rev. E {\bf 73}, 046105 (2006).

\bibitem{our}
A. Altland, A. De Martino, R. Egger, and B. Narozhny, 
preprint arXiv:1005.4662v1.

\bibitem{belzig}
M. Vanevic, Yu.V. Nazarov, and W. Belzig, Phys. Rev. Lett.
{\bf 99}, 076601 (2007).

\bibitem{abanov}
A.G. Abanov and D.A. Ivanov, Phys. Rev. Lett. {\bf 100}, 086602 (2008).

\bibitem{zhang}
J. Zhang, Y. Sherkunov, N. d'Ambrumenil, and B. Muzykantskii,
Phys. Rev. B {\bf 80}, 245308 (2009).

\bibitem{silva}
A. Silva, Phys. Rev. Lett. {\bf 101}, 120603 (2008).


\bibitem{gaspard2}
D. Andrieux, P. Gaspard, T. Monnai, and S. Tasaki, 
New J. Phys. {\bf 11}, 043014 (2009).

\bibitem{hanggi2a}
P. Talkner, E. Lutz, and P. H\"anggi,
 Phys. Rev. E {\bf 75}, 050102(R) (2007).

\bibitem{hanggi2b}
M. Campisi, P. Talkner, and P. H\"anggi,  
preprint arXiv:1003.1052v1.

\bibitem{hanggi2c}
M. Campisi, P. Talkner, and P. H\"anggi,  
preprint arXiv:1006.1542v1.

\bibitem{alex}
A. Altland and B.D. Simons, \textit{Condensed matter field theory},
2nd edition (Cambridge University Press, Cambridge, 2010). 

\bibitem{nazarov}
Yu.V. Nazarov and Ya.M. Blanter, \textit{Quantum Transport: Introduction
to Nanoscience} (Cambridge University Press, Cambridge, 2009).

\bibitem{kamenev_noneq}
A. Kamenev, in \textit{Nanophysics: Coherence and Transport}, Les Houches
session LXXXI, edited by H. Bouchiat,
Y. Gefen, S. Gu{\'e}ron, G.  Montambaux, and J. Dalibard
(Elsevier, New York, 2005).

\bibitem{keldysh}
A. Kamenev and A. Levchenko, Adv. Phys. {\bf 58}, 197 (2009).

\bibitem{aleiner}
I.L. Aleiner, P.W. Brouwer, and L.I. Glazman, 
Phys. Rep. {\bf 358}, 309 (2002).

\bibitem{schon}
G. Sch\"on and A.D. Zaikin, Phys. Rep. {\bf 198}, 237 (1990).

\bibitem{ourprl}
A. Altland and R. Egger, Phys. Rev. Lett. {\bf 102}, 026805 (2009).

\bibitem{kubo}
R. Kubo, K. Matsuo, and K. Kitahara, 
J. Stat. Phys. {\bf 9}, 51 (1973).

\bibitem{metiu}
K. Kitahara and H. Metiu, J. Stat. Phys. {\bf 15}, 141 (1976).

\bibitem{hanggi1}
P. H\"anggi, Z. Phys. B {\bf 31}, 407 (1978).

\bibitem{pilgram} 
S. Pilgram, A.N. Jordan, E.V. Sukhorukov, and M. B\"uttiker,
Phys. Rev. Lett. {\bf 90}, 206801 (2003).

\bibitem{tobiska}
J. Tobiska and Yu.V. Nazarov, Phys. Rev. B {\bf 72}, 235328 (2005).

\bibitem{fcs}
L.S. Levitov, H.-W. Lee, and G.B. Lesovik,
J. Math. Phys. {\bf 37}, 4845 (1996).

\bibitem{buttiker}
H. F\"orster and M. B\"uttiker, Phys. Rev. Lett. {\bf 101}, 136805 (2008).

\bibitem{saito1}
K. Saito and Y. Utsumi, Phys. Rev. B {\bf 78}, 115429 (2008).

\bibitem{saito2}
K. Saito and Y. Utsumi, Phys. Rev. B {\bf 79}, 235311 (2009).
 
\bibitem{ensslin}
S. Gustavsson, R. Leturcq, M. Studer, I. Shorubalko, T. Ihn, K. Ensslin,
D.C. Driscoll, and A.C. Gossard, Surf. Sci. Rep. {\bf 64}, 191 (2009).

\bibitem{fujisawa}
T. Fujisawa, T. Hayashi, R. Tomita, and Y. Hirayama, Science
{\bf 312}, 1634 (2006).

\bibitem{vanKampen}
N.G. Van Kampen, \textit{Stochastic Processes in Physics and Chemistry},
3rd edition (Elsevier, Amsterdam, 2007).

\bibitem{KolomeiskyFisher}
A. Kolomeisky and M.E.~Fisher, Annu. Rev. Phys. Chem. {\bf 58}, 675 (2007).

\bibitem{SchmiedlSeifert}
T.~Schmiedl and U.~Seifert, J. Chem. Phys. {\bf 126}, 044101 (2007).

\bibitem{MustonenLassig}
V.~Mustonen and M.~L\"assig, PNAS {\bf 107}, 4248 (2010).

\bibitem{zon1}
R. van Zon and E.G.D. Cohen, Phys. Rev. Lett.  {\bf 91}, 110601 (2003).

\bibitem{zon2}
R. van Zon, S. Ciliberto, and E.G.D. Cohen, 
Phys. Rev. Lett. {\bf 92}, 130601 (2004).

\bibitem{tc}
T. Taniguchi and E.G.D. Cohen, J. Stat. Phys. {\bf 130}, 633 (2007).

\bibitem{foot0}  
To keep the notation simple, we omit explicit time arguments 
whenever possible, $n_t\to n$, etc. 

\bibitem{naz}
Yu.V. Nazarov, Ann. Phys. (Leipzig) {\bf 16}, 720 (2007).

\bibitem{msr}
P.C. Martin, E.D. Siggia, and H.A. Rose,
Phys. Rev. A {\bf 8}, 423 (1973).

\bibitem{hkj}
H.K. Janssen, Z. Phys. B {\bf 23}, 377 (1976).

\bibitem{cd}
C. De Dominicis, J. Phys. (Paris), Colloq. {\bf 37}, 247 (1976).


\bibitem{onsager}
L. Onsager and S. Machlup, Phys. Rev.  {\bf 91}, 1505 (1953).

\bibitem{kurchan}
L.F. Cugliandolo, D.S. Dean, and J. Kurchan, 
Phys. Rev. Lett. {\bf 79}, 2168 (1997).

\bibitem{langevin2}
V.Y. Chernyak, M. Chertkov, and C. Jarzynski, J. Stat. Mech.  P08001 (2006).

\bibitem{sivan}
U. Sivan, Y. Imry, and A.G. Aronov, Europhys. Lett. {\bf 28}, 115 (1994). 

\bibitem{blanter}
Ya.M. Blanter, Phys. Rev. B {\bf 54}, 12807 (1996).

\bibitem{altshuler}
B.L. Altshuler, Y. Gefen, A. Kamenev, and L.S. Levitov,
Phys. Rev. Lett. {\bf 78}, 2803 (1997).

\bibitem{pilgram2}
S. Pilgram, 
Phys. Rev. B {\bf 69}, 115315 (2004).

\bibitem{kindermannpilgram}
M. Kindermann and S. Pilgram, 
Phys. Rev. B {\bf 69}, 155334 (2004).

\bibitem{hekkilanazarov}
T.T. Heikkil\"a and Y.V. Nazarov,
Phys. Rev. Lett.  {\bf 102}, 130605 (2009).

\bibitem{footeff} The temperatures $T_\nu^\ast$ in Eq.~\eqref{eq:2}
may still depend on the variable $n$.  In a quasi-stationary 
regime with $\partial_t n\approx 0$, we have $n\simeq n(V_\nu,T_\nu)$,
and a precise FR involving the effective
 temperatures $T_\nu^\ast$ can be stated.

\bibitem{footmore}
The slight mismatch between the slopes obtained from numerics and 
from the FR visible in Fig.~\ref{fig1} is 
 caused by the finite size of the window used for the numerical 
sampling of $Q$, which in turn is necessary to have
reasonable statistical efficiency.  

\end{thebibliography}
\end{document}